# New evidence for a nonclassical behavior of laser multimode light

M. Lebedev [1,2], A. Demenev [1], A. Parakhonsky [1]* and O. Misochko [1,2]

[1]Institute of Solid State Physics, Moscow region, Chernogolovka 142432, Russia;
[2]Moscow Institute of Physics and Technology, Moscow region, Dolgoprudny 141701, Russia

*alpar@issp.ac.ru

**Abstract:** In this work, we present new experimental evidence of a nonclassical behavior of a multimode Fabry-Perot (FP) semiconductor laser by the measurements of intensity correlation functions. Because of the multimode quantum state occurrence, instead of expected correlations between the intensities of the laser modes (a semiclassical theory), their anticorrelations were revealed.

## 1. Introduction

Our previous experimental results have showed strong photon correlations in a radiation of a multimode semiconductor laser [1]. The explanation of this effect could be given in a frame of semiclassical theory of light as well as in a frame of totally quantum mechanical picture, including the back action of photon detection on the emission of the subsequent photon. Trying to distinguish between these two possibilities we have performed triple intensity correlation measurements, which pointed on a quantum mechanical explanation [2, 3]. Here, we present a further proof of quantum character of a multimode FP semiconductor laser radiation based on a novel experimental scheme. It contains four single photon avalanche diodes (SPAD), which allow recording double correlators of three kinds: the multimode autocorrelation functions, the single mode autocorrelation functions and all-to-one cross-correlation functions. The idea of our measurements is a careful comparison of individual mode autocorrelation functions. Surprisingly, we observe an anticorrelation between all-to-one intensities instead of their correlation predicted by a semiclassical theory.

## 2. Experiments and Methods

The experimental setup is schematically shown in Figure 1. Radiation of the multimode semiconductor FP laser (model FPL-852+/−2 nm, Nolatech) [2] is splitted by a beamsplitter (a parallel-sided glass plate) into two beams. One beam is attenuated by a neutral density filter (NF) and focused into a multimode fiber (f1), after which it is distributed between two SPADs (D1, D2) by means of a fiber-Y-splitter (FBS1). Another beam is directed onto the entrance slit of a triple grating spectrometer Ramanor U1000. The output beam is then focused into a multimode fiber (f2), after which it is distributed between the photodiodes D3 and D4 by means of the fiber-Y-splitter FBS2.

The multimode spectrum of the laser operating in its standard regime with a total power of 30 mW is shown in the inset of Figure. 1. The spectrum consists of a large number (approximately 35) of longitudinal modes. All avalanche photodiodes are operated in the photon counting regime. Using the high resolution triple spectrometer Ramanor U1000 we can select one of the modes from the spectrum shown in the inset of Figure 1. This allows us to measure the second-order intensity correlation functions of three kinds: a correlation function of the multimode radiation, a single mode correlation function and a cross-correlation function between all modes and a single one (see below). The slits used in the spectrometer were only some microns wide in order to achieve the best possible spectral resolution, as well as to reduce the light intensity.

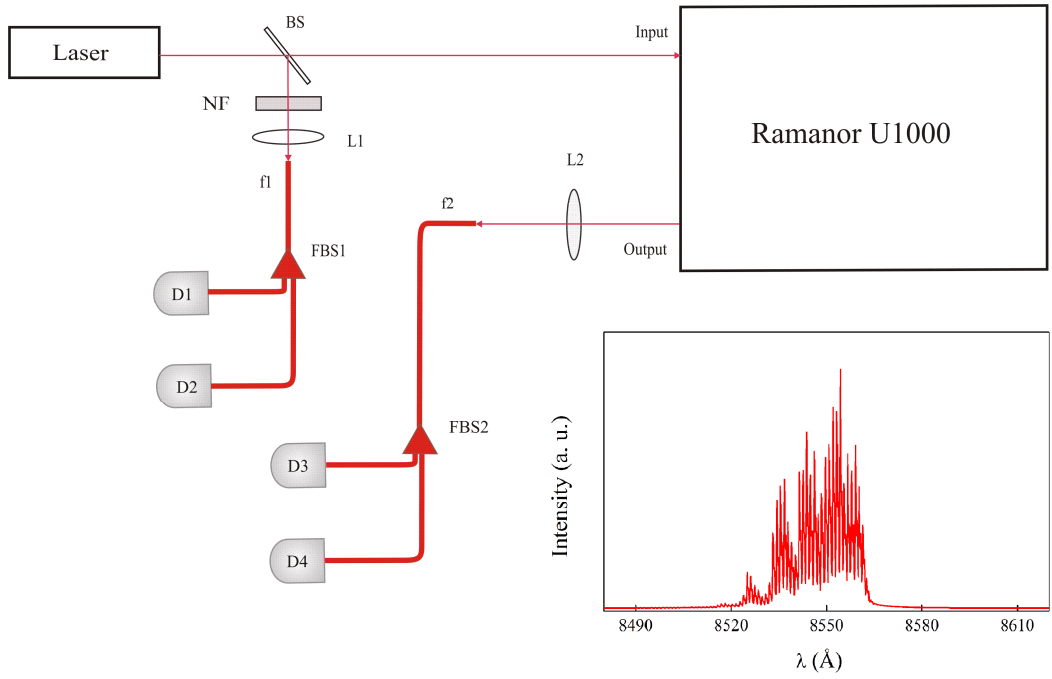

**Figure 1.** Schematic of the experimental setup. The laser is the multimode semiconductor FP laser [2]. NF is a neutral density filter. BS is a beamsplitter (a parallel-sided glass plate). FBS1 and FBS2 are fiber-Y-splitters. f1 and f2 are multimode optical fibers. D1, D2, D3 and D4 are single photon detectors (silicon avalanche photodiodes) or SPADs. L1 and L2 are lenses. Ramanor U1000 is a triple grating spectrometer. The multimode spectrum of the laser is shown in the right lower corner of the figure.

## 3. Results

The second-order correlation function $g_\forall^{(2)}$ for the multimode radiation is

$$g_\forall^{(2)}(\tau) = \frac{\langle I_\forall(0) I_\forall(\tau) \rangle}{\langle I_\forall \rangle^2}, \quad (1)$$

where $I_\forall(0)$ and $I_\forall(\tau)$ are the intensities measured with detectors D1 and D2 respectively.

This function shows the absence of any significant correlations as it is expected for a coherent state (see Figure 2).

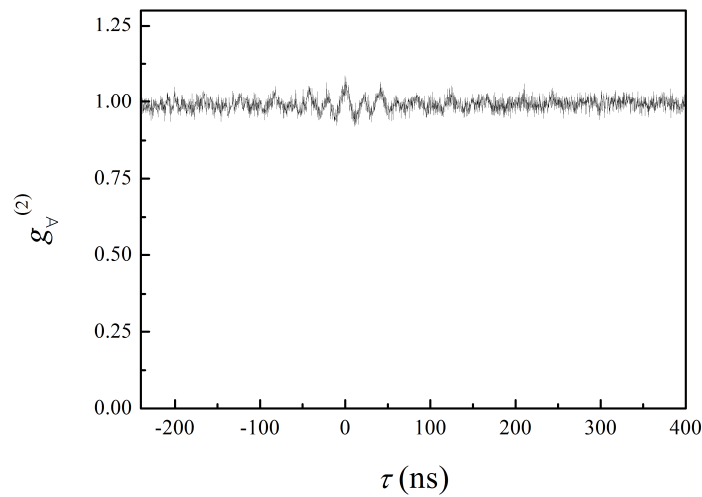

**Figure 2.** The second-order intensity correlation function of the multimode FP laser radiation.

A multimode continuous-wave laser is often thought to generate a state which is a product:

$$|\Psi\rangle = C|\alpha_1\rangle...|\alpha_n\rangle, \quad (2)$$

where $C$ is the normalization constant and $|\alpha_1\rangle ... |\alpha_n\rangle$ are coherent states of the individual modes, thus the absence of correlations is quite natural.

As above mentioned, the total number of intense longitudinal modes in the spectrum of the laser working at its nominal optical power of 30 mW was about 35. We have measured autocorrelation functions for 25 modes (see Figure 3). The second-order intensity correlation functions $g_i^{(2)}$ of a desired single mode is

$$g_i^{(2)}(\tau) = \frac{\langle I_i(0)I_i(\tau)\rangle}{\langle I_i\rangle^2}, \quad (3)$$

where $I_i(0)$ and $I_i(\tau)$ are the intensities of a single mode measured with the detectors D3 and D4.

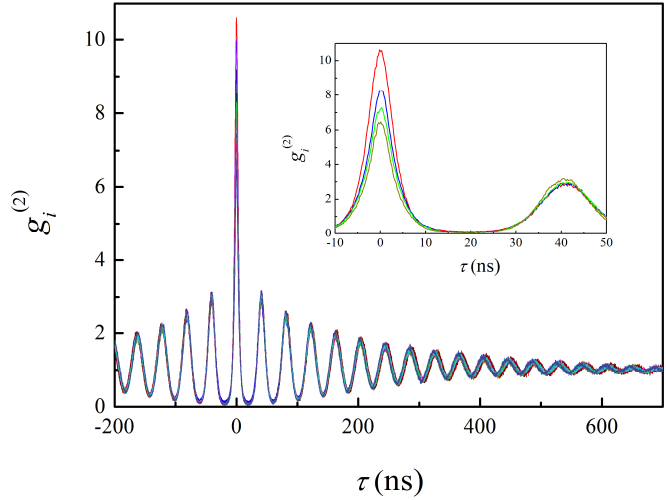

**Figure 3.** The intensity correlation functions of single modes. All the functions are almost identical. The inset shows four selected functions close to the first minimum corresponding to four single modes (red curve: λ=8510.7 Å, blue curve: λ=8533.5 Å, green curve: λ=8535.0 Å and dark yellow curve: λ=8538.0 Å).

All the modes are almost identical regarding their autocorrelation functions. The visibility of these functions is extremely high, with the level very close to 100% (the counting rate in the first minimum of the autocorrelation function was only (2-5) counts/1000 s). It can be also observed from Figure 3 (in the inset) that the maximum of the function is inversely proportional to the wavelength (the values are given in the figure caption). These functions show a strong photon bunching at zero time delay followed by oscillations with a period of about 40 ns and a damping constant of about 190 ns. The only difference between autocorrelation functions of individual modes is the height of the zero-delay peak. This value varies between 4.5 and 11 and shows also some instability if the measurements of single mode autocorrelation functions are repeated for several times. We attribute this instability to intensity fluctuations caused by spatial interference effects at the entrance slit of the spectrometer.

The cross-correlation function between all modes and a single one can be expressed as

$$g_{i,\forall}^{(2)}(\tau) = \frac{\langle I_i(0)I_\forall(\tau)\rangle}{\langle I_i\rangle\langle I_\forall\rangle}. \quad (4)$$

where $I_\forall(\tau)$ and $I_i(0)$ are the intensities of all modes and a single one measured with detectors D1 and D3.

Surprisingly, we found that the maximum absolute value of correlations actually occurs at the time delay of 40 ns which is roughly comparable with a time delay corresponding to the propagation time of light from the beamsplitter through the spectrometer and the output fiber to the detector D3. However, we observe an anti-correlation between the intensities $I_1$ and $I_3$ (measured by means of detectors D1 and D3, correspondingly) instead of a correlation at this time delay (see Figure 4).

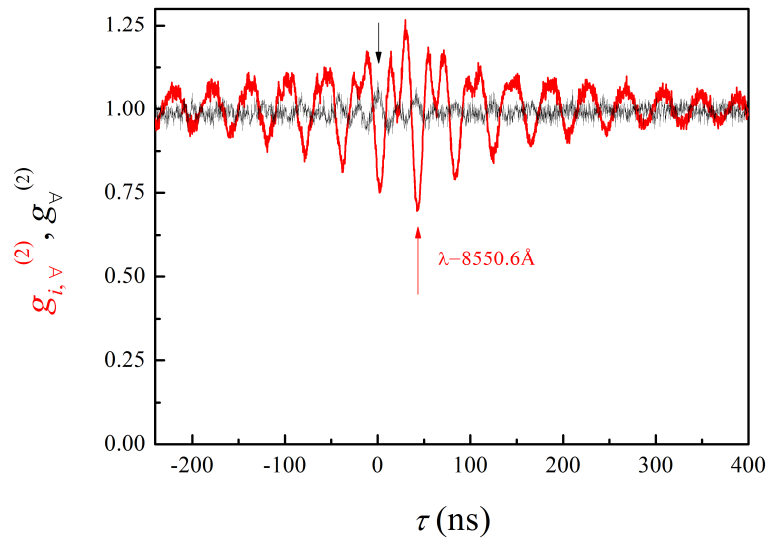

**Figure 4.** The intensity cross-correlation function (red curve) measured between all modes and a single mode (λ=8550.6 Å). The intensity correlation function of the multimode radiation is shown once again for the comparison (grey curve).

### 4. Discussion

It is clear that our results obviously contradict the formula (2), showing that individual modes are not in coherent states. A high-visibility intensity correlation function may be a signature of pulsed-laser operation, and such a pulse regime of a continuous wave multimode semiconductor laser was indeed observed experimentally [4]. In this case, continuous-wave multimode laser operation is a mixture of pulse trains of individual modes which phases are chosen so that the integrated intensity is kept constant. In our case, the intensity correlation functions measured in a wide time scale showed no regular pulse trains repeating intensity fluctuations observed at zero-time delay. Nevertheless, we believe that such a pulse regime can indeed occur but this is insufficient to explain the strong correlations observed between individual photons. In other words, the strong photon bunching should look like a pulse train in a high power laser beam but the reverse is not automatically true: a high power laser pulse train does not demand strong photon bunching. For example, if we take a chaotic light source like a neon lamp (Ne) and modulate its radiation with an external modulator then intensity correlation experiments reveal the modulation envelope but photons remain uncorrelated within the envelope. On the contrary, a quantum correlated state implies correlations between the individual photon events.

Within the framework of a semiclassical theory, one can regard photon detection with SPADs as appropriate stationary random processes of the intensity fluctuations. We can see that functions $g_\forall^{(2)}$ and $g_i^{(2)}$ have their maxima at zero-time delay as it should be for a stationary random process. This means that the maximum intensity correlation is observed at the wave front of the laser radiation (both for a pair of detectors D1 - D2 and also for D3 - D4). Then, one could expect that a maximal correlation should be observed between detectors D1 and D3 at a time delay corresponding to the propagation time from the beamsplitter to the detector D3. Namely, while the wave front measured with detector D1 arrives to detector D3. In other words, a short light pulse will give a maximum value of the autocorrelation function at zero time delay measured with detectors D1 and D2 as well as measured with detectors D3 and D4. It would seem the same is true of detectors D1 and D3: such a pulse will give a maximum value of the cross-correlation function $g_{i,\forall}^{(2)}$ at a time delay corresponding to the propagation time of the pulse through the spectrometer and the output fiber to detector D3.

It should be noted that the cross-correlation curve shown in Figure 4 has a minimum at a point of the greatest value of correlations. The comparison of the curves clearly shows that the maxima of the one curve coincide with the minima of the other and vice-versa.

For the avoidance of any unexpected apparatus effects, we measured also the intensity correlation functions where the first detector (D1) was illuminated with a Ne lamp, while a single mode laser light was registered by the second detector (D3) (see Figure 5). In addition, the first detector was illuminated with a multimode laser light, while the second was used to register Ne light. In both cases no correlation occurrences were revealed.

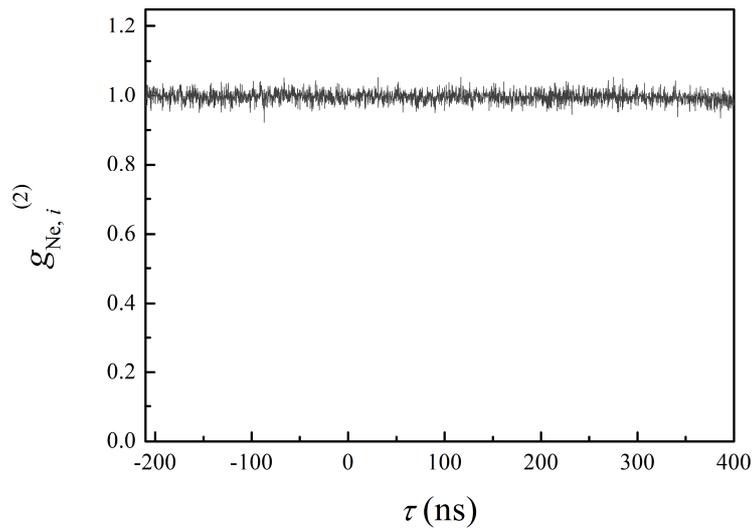

**Figure 5.** The intensity cross-correlation function between a Ne lamp and a single mode of the laser studied.

## 5. Conclusions

In conclusion, the measurements of second-order $g^{(2)}$ intensity correlation functions of a multimode semiconductor Fabry-Perot laser have shown that its individual modes are not in coherent states, even though they have identical autocorrelation functions. These functions reveal a strong photon bunching followed by pronounced oscillations. The antibunching effect in cross-correlation function measurements between the radiation of individual mode and the total radiation of the laser cannot be explained by the semiclassical theory of periodic pulse trains generated by a multimode semiconductor laser. Obviously, the quantum-mechanical model is needed to describe the generation of multimode light in a FP semiconductor laser.

**Funding:** This research was funded by the Russian Foundation for Basic Research, grant number 20-02-00231.